\documentclass[aps,prd,reprint,superscriptaddress,preprintnumbers,nofootinbib]{revtex4-1}

\usepackage{amsmath}
\usepackage{amsfonts} 
\usepackage{amssymb}
\usepackage{graphicx}
\usepackage{hyperref}
\usepackage{tensor}
\usepackage{siunitx}
\usepackage{physics}
\usepackage{cancel}
\usepackage{slashed}
\newcommand{\pa}{\partial}
\usepackage{appendix}
\newcommand{\be}{\begin{equation}}
\newcommand{\ee}{\end{equation}}

\newcommand{\beq}{\begin{equation}}  \newcommand{\eeq}{\end{equation}}
\newcommand{\beqn}{\begin{eqnarray}}
 \newcommand{\eeqn}{\end{eqnarray}}
\newcommand{\bal}{\begin{aligned}}   \newcommand{\eal}{\end{aligned}}
\newcommand{\bea}{\begin{eqnarray}}  \newcommand{\eea}{\end{eqnarray}}

\newcommand{\rd}{\mathcal}

\hypersetup{colorlinks=true,
			urlcolor=purple,
			citecolor=blue,
			linkcolor=magenta,}

\begin{document}

\title{\Large  Infinite Black Hole Entropies at Infinite Distances \\and Tower of States}
\preprint{LMU--ASC 54/19}
\preprint{IPhT-T19/163}
\preprint{CPHT096.122019}
\preprint{MPP-2019-257}
\preprint{DESY 19-226}

{~}

\author{\vspace{0.5cm}
\large Quentin~Bonnefoy\vspace{0.5cm}}
 \affiliation{DESY, Notkestra{\ss}e 85, 22607 Hamburg, Germany\\[1.5ex]}
\author{\large Luca~Ciambelli}
 \affiliation{Universit\' e Libre de Bruxelles and International Solvay Institutes, ULB-Campus Plaine CP231, B-1050 Brussels, Belgium\\[1.5ex]}
\author{\large Dieter~L\"ust}
 \affiliation{Arnold-Sommerfeld-Center for Theoretical Physics, Ludwig-Maximilians-Universit\"at, 80333 M\"unchen, Germany\\[1.5ex]}
\affiliation{Max-Planck-Institut f\"ur Physik (Werner-Heisenberg-Institut),
             F\"ohringer Ring 6,
             80805, M\"unchen, Germany\\[1.5ex]}
\author{\large Severin~L\"ust}
 \affiliation{CPHT, CNRS, Ecole Polytechnique, IP Paris, F-91128 Palaiseau, France\\[1.5ex]}
  \affiliation{Institut de Physique Th\'eorique, Universit\'e Paris Saclay, CEA, CNRS, F-91191 Gif sur Yvette, France}


\begin{abstract}
\vspace{1.0cm}
\noindent
The aim of this paper is to elucidate a close connection between the black hole area law and the infinite distance conjecture 
in the context of the swampland.
We consider families of black hole geometries, parametrized by their event horizon areas or by the values of their entropies, and show that the infinite entropy limit is
always at infinite distance in the space of black hole geometries. 
It then follows from the infinite distance conjecture that there must be a tower of  states in the infinite 
entropy limit, and that ignoring these towers on the horizon of the black hole would invalidate the effective theory when the entropy becomes large. We call this the black hole entropy distance conjecture.
We then study two candidates for the tower of states. The first are the Kaluza-Klein modes of the internal geometry of extremal ${\cal N}=2$ black holes in string theory, whose masses on the horizon are fixed by the ${\cal N}=2$ attractor formalism, and given in terms of the black hole charges similarly to the entropy. However, we observe that it is possible to decouple their masses from the entropy, so that they cannot 
generically  play the role of the tower. We thus consider a second kind of states: inspired by N-portrait quantum models of non-extremal black holes, we argue that the Goldstone-like modes that interpolate among the black hole microstates behave like the expected light tower of states. 
 \vspace{1cm}
\end{abstract}

\maketitle

\section{Introduction}

In this paper we want to generalize the infinite distance conjecture of the swampland scenario \cite{Vafa:2005ui,Ooguri:2006in,Palti:2019pca} to black hole geometries with horizon.
The famous area law of Bekenstein and Hawking \cite{Bekenstein:1973ur,Hawking:1974sw} states that the entropy ${\cal S}$ of a particular space-time  geometry 
is proportional to the surface area $A$ of its event horizon:
\begin{equation}\label{bhrel}
{\cal S}={1\over 4}{A\over  L_p^2}\, ,
\end{equation}
where $L_p$ is the Planck length. Viewing this relation from the perspective of statistical mechanics, it means that the number of microstates of a space-time geometry exponentially grows
with the area of the event horizon. The agreement between the macroscopic and microscopic entropies was shown first for supersymmetric black holes in string theory 
\cite{Strominger:1996sh},
and subsequently
investigated for several space-time geometries in string theory and in supergravity. 
For  the Schwarzschild geometry the relation between the area law and black hole microstates is less understood, but progress
towards its understanding was made by investigating the so-called soft hairs of black holes \cite{Hawking:2015qqa,Hawking:2016msc,Averin:2016ybl,Averin:2016hhm,Hawking:2016sgy,Haco:2018ske}. 
Finally, for quantum field theories there is close relation between entropy and the number of quantum degrees of freedom
via renormalization group flow and the c-theorem 
\cite{Zamolodchikov:1986gt,Komargodski:2011vj}.

\vskip0.3cm
A priori unrelated to the entropy/area law in quantum gravity, 
 it is argued in the context of the swampland approach \cite{Vafa:2005ui,Ooguri:2006in,Palti:2019pca} that
large distances in the field range of effective field theories 
lead to a tower of states
when coupled to quantum gravity.
This observation has led to the infinite distance conjecture stating that 
if, in quantum gravity, one tries to increase the range of some field $\phi$
beyond the Planck range some new light
states emerge, invalidating the effective
field theory.
Specifically  the masses  of these additional states  exponentially decrease  as functions of the distance in field space,  
\be
m \sim M_p e^{-c |\Delta(\phi)| }\;,
\label{gendisca}
\ee
where $c \sim {\cal O}(1)$.
 Hence for large field distances with $|\Delta(\phi)|\rightarrow\infty$, we get in quantum gravity a massless tower of states, which invalidates the effective field theory description.

\vskip0.3cm
The infinite
distance conjecture 
was discussed  in string theory in many interesting instances 
\cite{Klaewer:2016kiy,Heidenreich:2017sim,Andriolo:2018lvp,Grimm:2018ohb,Heidenreich:2018kpg,Blumenhagen:2018nts,Blumenhagen:2018hsh,Lee:2018urn,Lee:2018spm,Grimm:2018cpv,Klaewer:2018yxi,Bonnefoy:2018tcp,Corvilain:2018lgw,Lee:2019tst,Joshi:2019nzi,Marchesano:2019ifh,Font:2019cxq,Lee:2019xtm,Bachas:2019rfq,Erkinger:2019umg,Lust:2019zwm,Agrawal:2019dlm,Gomez:2019ltc,Marchesano:2019hfb,Bedroya:2019snp,Kehagias:2019akr,Lee:2019wij,Grimm:2019bey,Geng:2019zsx,Grimm:2019ixq,Blumenhagen:2019vgj,Kehagias:2019iem,Anchordoqui:2019amx,Font:2019uva,Scalisi:2019gfv}.
In the context of string compactifications,
 $\phi$ is labelling a family of  internal compact spaces, corresponding e.g.~to a size modulus $R$ of an internal Calabi-Yau manifold.
 Then in the limit where some internal cycles either become very large or shrink to zero size, generically two kind of towers emerge:
 (i) an infinite tower of Kaluza-Klein (or winding states), which scale as $m_\text{KK}\sim n/R$ (or $m_\text{wind.}\sim nR$ in string units), where $n$ is some integer and $R$ is the size of the internal cycle. The associated distance is given as 
 \begin{equation}\label{r}
 \Delta(R)\sim |\log R|\, .
 \end{equation}
 (ii) Second, one can get also tensionless strings, and the associated tower becomes
 light in the limit of weak string coupling and scales as $m_{\rm string}\sim \sqrt{n} g_s$, where $g_s$ is the string coupling constant. Here the distance in the string coupling constant is 
 \begin{equation}\label{gs}
 \Delta(g_s)\sim |\log g_s|\, .
 \end{equation}
 In fact, as  advocated in \cite{Lee:2019wij}, these two kinds of states are the only possible physical towers in the context of string compactifications.

 \vskip0.3cm
 The infinite distance conjecture can be also generalized to effective gravity theories, namely to the  variation  of the background metric itself.
 Here we
 consider gravity with a family of background metrics $g_{\mu\nu}(\tau)$, which are labeled by some parameter $\tau$.
 Then the generalized distance conjecture  \cite{Lust:2019zwm} implies  that, if the associated distance $\Delta(\tau)$  in the space of background metrics becomes infinite, always a light tower
 of states has to emerge. This generalized distance conjecture was applied in \cite{Lust:2019zwm} to AdS space-times, which are labeled by the values of their (negative) cosmological constants $\Lambda$.
 Then the AdS distance conjecture (ADC) states that  the corresponding infinite tower of states scales as 
$
m_\text{AdS} \sim |\Lambda|^{\alpha}  $,
 with $\alpha=1/2$ for the strong ADC.

\vskip0.3cm
When considering the variations  of the background metric,
  it was also argued   \cite{Kehagias:2019akr}  that in gravity the Ricci-flow \cite{hamilton} and also generalized gradient flows \cite{Perelman:2006un}
provide a refined
criterion for the distance conjecture and its connection to the swampland: namely
there exists an infinite tower of physical states which become massless when following the Ricci-flow towards a fixed point at infinite distance.
The corresponding ``Ricci distance"  is given by the scalar curvature ${\cal R}$ of the background metric: 
\begin{equation}\label{riccidistance}
\Delta({\cal R})\sim |\log {\cal R}|\, .
\end{equation}
This equation can be viewed as the generalization of the distance $\Delta(R)$ in \eqref{r}. 
In addition,  entropy functionals, 
which are related to the more general gradient flow equations, provide a sensible definition 
of the generalized distance in the space of background fields.
Considering the combined dilaton-metric flow, the corresponding grading flow is derived from the entropy functional ${\cal F}$, which then provides a good definition for the distance in the combined 
metric-dilaton field space
\cite{Kehagias:2019akr}:
\begin{equation}
\Delta({\cal F})\sim|\log {\cal F}|\, .
\end{equation}
For example, if one considers the flow of the string coupling constant $g_s(\tau)$, one
sees that the distance $\Delta({\cal F})$ agrees with the distance in \eqref{gs} and confirms that at least in the space of metrics and string coupling constants these two distances and their associated two towers of states are the only ones
 appearing in the context of string compactifications.

\vskip0.3cm
In this paper 
we consider effective gravity theories with a family of black hole metrics $g_{\mu\nu}(r_S)$ with horizons of size $r_S$.
As we will discuss, the limit of 
infinite horizon $r_S\rightarrow\infty$ 
is at infinite distance in 
 the space of metrics and
the associated horizon distance  can be always expressed as 
\begin{equation}
 \Delta({r_S})\sim\log r_S\, .
 \end{equation}
 Using the Bekenstein-Hawking relation (\ref{bhrel}) between the size of the horizon and the entropy ${\cal S}$ of the horizon geometry, namely
 \begin{equation}
 {\cal S}\sim r_S^2\, ,
 \end{equation}
 it follows that the distance can be expressed in terms of the entropy as
 \begin{equation}
 \Delta({\cal S})\sim\log {\cal S}\, .\label{entropydistance}
 \end{equation}
Now  applying the infinite distance conjecture,   there must be a corresponding mass scale with a tower of ``states", whose
 masses in Planck units,  according to \eqref{gendisca}, are given as
\be
m_{\cal S} \sim {\cal S}^{-c}\;,
\label{gendisc}
\ee
where $c$ is a positive constant. 
We call this the {\sl black hole entropy distance conjecture}, or in short BHEDC. 
For horizons larger than the Planck distance and correspondingly for large entropies this mass scale becomes very tiny.
This means that the effective field theory of any macroscopic black hole must contain a very large number of almost massless modes.
For $c=1/2$, the black hole distance conjecture looks very similar to the strong ADC, and therefore this case is called strong BHEDC.

\vskip0.3cm
{Black holes are best understood from string compactifications when they are extremal, so we first look at extremal ${\cal N}=2$ black holes and study as candidates for a tower of states
the Kaluza-Klein modes from the internal geometry. Here the attractor equations of supersymmetric extremal black holes \cite{Ferrara:1995ih,Ferrara:1996dd} imply that large horizons and large entropies follow from large values of internal moduli, i.e.~light Kaluza-Klein modes on the horizon. However, we show that the reverse statement is not true, and that the Kaluza-Klein modes can be kept heavy while going to the infinite entropy limit. Thus, those modes cannot  play the generic role of the tower.}


\vskip0.3cm
{We consider a second option, namely we }address the problem directly on the black hole solution, without relying on higher-dimensional constructions.  {In this setup, the infinite distance conjecture 
 infers a tower prediction originating from the gravitational modes of the lower-dimensional effective gravity theory. We consider non-extremal black holes which are less understood from the point of view of string compactifications, so we rely instead on the microscopic black hole N-portrait model of \cite{portrait,portrait1,Dvali:2013eja,portrait3,portrait4,gold}.} As we will argue, for non-extremal black holes, like the Schwarzschild black hole, the mass scale $m_{\cal S}$ belongs to a new kind of tower of ``states": 
these states are not given in terms of Kaluza-Klein from the higher-dimensional theory, but this tower 
is related  to the 
carriers of the information, namely they are closely related to the almost gapless modes 
that describe the degeneracies among the microstates  of the black hole geometry. 
 These modes were extensively discussed in the N-portrait model. 
 For finite horizons and finite entropies, the ``masses", or better say the frequencies of these modes are finite. On the other hand, in the infinite horizon and infinite entropy limit
 the frequencies of the modes connecting the infinite microstates become zero. 
This tower behaviour  of non-extremal black holes is closely related to  the soft hair proposal 
 \cite{Hawking:2015qqa,Hawking:2016msc,Averin:2016ybl,Averin:2016hhm,Hawking:2016sgy,Haco:2018ske}
  for the black hole entropy.
 The limit $r_S\rightarrow\infty$ corresponds to flat  Minkowski space, where the horizon approaches null
 infinity of Minkowski space.
 In this limit the entropy of flat space is infinite
 \cite{Dvali:2015rea,Averin:2016ybl}
and the corresponding microstates of Minkowski space can be viewed as the infinite number of soft gravitons, which are associated with the infinite number of BMS transformations at null infinity.
From the swampland context, the emergence of  this massless tower of modes  in the infinite entropy limit of non-extremal black holes means that the
 effective horizon field theory must contain these states. If this is not the case, then the effective theory is in the swampland.

\vskip0.3cm
The paper is organized as follows. {In section \ref{sec3}, we briefly review the reasoning behind the ADC and the generalized distance conjecture of \cite{Lust:2019zwm}, together with the definition of a certain geometric distance functional in the space of metrics. We also repeat the statement of the BHEDC.
In sections \ref{section:oneparameter} - \ref{sec5}, we extend the distance conjecture to black hole geometries. We show that the limit of large horizon, or equivalently the limit of large entropy, is at infinite distance in the space of black hole metrics. In  section \ref{section:n=2} we discuss the large entropy limit of ${\cal N}=2$ extremal black holes in string theory.
As mentioned before, via the attractor equations, this limit is correlated to a limit of large moduli fields of the internal compact space. However, although large moduli imply large entropy, we show that the reverse is not true and that the Kaluza-Klein states of the internal manifold do not provide a satisfying tower of states. The tower of light modes for non-extremal black holes in the N-portrait model is then discussed in section \ref{sec6}. We present some conclusions in section \ref{sec7}. We exile to appendix \ref{appdiffeos} some remarks on the properties of the geometric distance formula used here under diffeomorphisms that depend on the parameter $\tau$ along the path, and we also discuss an alternative definition of the distance functional in terms of the scalar curvature of spacetime. In appendix \ref{appvolume} we compare the formula used in this paper to the analogue without regulating volume factor, which appeared often in the literature. Finally in appendix \ref{appC}, we present some further details about the quasi Goldstone modes, their energies
and their relation to the black hole microstates.}

\section{The ADC, the geometric distance formula and the BHEDC}\label{sec3}

{In \cite{Lust:2019zwm} it was argued that the infinite distance in the space of AdS space-times always comes with an exponentially light tower of states. This behaviour was called the AdS distance conjecture (ADC), and the precise statement is that, when the cosmological constant $\Lambda\to0$, a tower of states becomes massless with the following behaviour:
\be
m\propto \abs{\Lambda}^c \ ,
\ee
where $c={\cal O}(1)$. This generalizes the usual distance conjecture \cite{Ooguri:2006in}: when a background metric travels a distance $\Delta(g)$, a tower of states becomes exponentially light,
\be
m\propto e^{-c\Delta(g)} \ .
\ee
The matching with the ADC comes from the fact that the distance for AdS is logarithmic in the cosmological constant. This can be seen by introducing a time-dependent fluctuation of the cosmological constant, $\Lambda\to\Lambda+\delta\Lambda(t)$, and computing the kinetic term of the fluctuation from the Einstein-Hilbert action:
\be
R\supset -\frac{(\partial \delta\Lambda)^2}{\Lambda^2} \ ,
\label{ccKineticTerm}
\ee
where we dropped an irrelevant dimension-dependent constant factor, and where the $\partial$ are contracted with the undeformed AdS metric. This is easily seen in global coordinates where the ($d$-dimensional) AdS metric reads
\be
\text ds^2=\frac{(1-d)(d-2)}{2\Lambda}(-\cosh^2(\rho) \, \text dt^2+\text d\rho^2+\sinh^2(\rho)\, \text d\Omega_{d-2}^2) \ .
\label{AdsGlobalCoos}
\ee
Then, changing the cosmological constant amounts to performing a Weyl rescaling of the metric, leading to the kinetic term \eqref{ccKineticTerm}.}

{The ADC is supported by several string theory examples, for instance by the AdS$_5\times$S$^5$ vacuum of string theory, where the volume of the S$^5$ goes to infinity when the cosmological constant goes to zero.
So in most of the explored cases the ADC is closely related to the observation that the $d$-dimensional AdS length scale cannot be decoupled from
the internal length scale of the additional compact space, which means that
 the masses of the Kaluza-Klein modes follow the behavior predicted by the ADC.}

{In what follows, we will extend the ADC to black hole space-times.} In order to analyze the behaviour of towers of states when we scan over physically different space-times, we first need to define a distance which measures the length of the geodesic path corresponding to this scan. In this paper, we use the geometric distance formula \cite{DeWitt}, 
\be
\bal
\Delta_g &=  c \int_{\tau_i}^{\tau_f}  \left( \frac{1}{V_M} \int_M \sqrt{g} g^{MN} g^{OP} \frac{\partial g_{MO}}{\partial \tau} \frac{\partial g_{NP}}{\partial \tau}   \right)^{\frac12}\text d\tau\\
&= c \int_{\tau_i}^{\tau_f}  \left( \frac{1}{V_M} \int_M \sqrt{g} \tr[\left(g^{-1}\frac{\pa g}{\pa \tau}\right)^2] \right)^{\frac12}\text d\tau\;,
\label{dismetr}
\eal
\ee
where $g$ is the metric on the manifold $M$, $V_M=\int_M\sqrt{g}$ is the volume of the latter, $\tau$ is a parameter along the path between $g(\tau_i)$ and $g(\tau_f)$ and $c \sim {\cal O}(1)$. This distance was used in \cite{Lust:2019zwm}
to formulate the AdS distance conjecture\footnote{{Strictly speaking, the use of this distance was only justified in \cite{Lust:2019zwm} for transverse traceless metric fluctuations. However, as we will soon see, it still captures the behaviour predicted by the ADC, as well as other behaviours to be discussed later. For this reason, we take \eqref{dismetr} as a distance which describes the relevant physics.}} and it leads to the same result
as applying the above mentioned Weyl rescalings. 

{A comment is in order at this point: in appendix \ref{appdiffeos}, we argue that the geometric distance functional is not covariant under diffeomorphisms involving $\tau$, the parameter labeling the family of metrics, so that particular care should be taken when computing the distance. However in the following, we will choose the prescription which was already successfully applied in the context of the ADC. 
Further evidence for the correctness of this prescription is provided by an alternative definition of the geometric distance between different space-time geometries, discussed in appendix \ref{covdisapp}.}

\subsection{Geodesic equation}

The shortest path between two space-time metrics can be identified by extremizing $\Delta_g$. This leads to the geodesic equation\footnote{Note that the geodesic equation is modified with respect to the case of a metric distance without the volume normalization, as in e.g. \cite{gilmedrano1992riemannian}. See also appendix \ref{appvolume} for details.}
\be
\bal
\ddot{g}=&\ \dot{g}g^{-1}\dot{g}+\tfrac14 \tr[(g^{-1}\dot{g})^2]g-\tfrac 12\tr[g^{-1}\dot{g}]\dot{g}\\
&-\tfrac14\langle\tr[(g^{-1}\dot{g})^2]\rangle g+\tfrac12\langle\tr[g^{-1}\dot{g}]\rangle\dot{g} \ ,
\eal
\ee
where
\be
\langle X\rangle=\frac{\int_M\sqrt{g}\ X}{V_M}
\ee
and a dot indicates differentiation with respect to the proper time\footnote{The geodesic equation holds for more general affine parameters $\lambda$, but in order to limit the number of unfixed constants in the solutions to be discussed later, we always express them in terms of the proper time. For instance, in \eqref{firstGeodesicSol}, the power $\sqrt{d}^{-1}$ in the exponential is fixed when one imposes that the affine parameter $\lambda$ is the proper time as defined in \eqref{properTime}.} $\lambda$, defined as
\be
\text d\lambda=\text d\tau\left(\frac{1}{V_M}\int_M \sqrt{g}\tr[\left(g^{-1}\frac{\partial g}{\partial \tau}\right)^2]\right)^{1/2} \ .
\label{properTime}
\ee
Defining
\be
f= g^{-1}\dot{g} \ ,
\label{ourF}
\ee
the geodesic equation can be concisely written as
\be
\dot{f}=\tfrac14\tr(f^2)\text{Id}-\tfrac12 \tr(f)f-\tfrac14\langle\tr(f^2)\rangle\text{Id}+\tfrac12\langle\tr(f)\rangle f\ .
\label{geodesicsVolume}
\ee

\subsection{First examples}\label{section:generalExamples}

We now want to solve \eqref{geodesicsVolume} to define geodesic paths between physically different space-times. Then, using \eqref{properTime}, the distance is given in terms of the proper time variation:
\be
\Delta_g =c\int_{\tau_i}^{\tau_f}\text d\lambda=c(\lambda_f-\lambda_i)\;.
\ee
Important examples of solutions to the geodesic equation \eqref{geodesicsVolume} are the metrics with a constant parameter $\alpha$ entering as an overall factor:
\be
\text ds^2=g_{\mu\nu}(x,\alpha)\text dx^\mu \text dx^\nu \text{ with }g_{\mu\nu}(x,\alpha)=\alpha \tilde g_{\mu\nu}(x) \ .
\label{conformalAnsatz}
\ee
Upgrading $\alpha \rightarrow \alpha(\lambda)$, and using \eqref{conformalAnsatz} as an ansatz, it is straightforward to see that the geodesic equation is solved for
\be
\alpha(\lambda)=\alpha_0\, e^{\lambda/\sqrt{d}} \,,
\label{firstGeodesicSol}
\ee
where $d$ is the dimension of $M$.
Therefore, the geodesic distance between two space-times of the form \eqref{conformalAnsatz} with conformal factors $\alpha_i$ and $\alpha_f$ is given by
\be
\Delta_g = c\sqrt{d}\log(\frac{\alpha_f}{\alpha_i})\;.
\label{distanceConformal}
\ee
{This discussion applies to $d$-dimensional AdS spaces, whose metric in global coordinates is given in \eqref{AdsGlobalCoos}, so that using \eqref{distanceConformal}, one shows that the distance computed with \eqref{dismetr} is indeed logarithmic with respect to the cosmological constant, which is nothing but the ADC.}

\subsection{The Black Hole Entropy Distance Conjecture}\label{section:BHEDC}

{We will now turn to the study of black hole space-times. Let us anticipate and state the result: the distance will always be of the form
\begin{equation}
 \Delta({r_S})\sim\log r_S\, ,
 \end{equation}
 for black hole horizons of size $r_S$. Using the Bekenstein-Hawking relation (\ref{bhrel}), we immediately get that the distance can be expressed in terms of the entropy as
 \begin{equation}
 \Delta({\cal S})\sim\log {\cal S}\, .
\label{SDistance}
 \end{equation}
The statement of the BHEDC now amounts to applying the infinite distance conjecture to \eqref{SDistance}: there must be a tower of ``states", whose
 masses in Planck units are given by
\be
m_{\cal S} \sim {\cal S}^{-c}\;.
\label{BHEDCstates}
\ee
In what follows, we will first establish the statement \eqref{SDistance} and later discuss the interpretation of the conjectured tower of states.}

\section{One-parameter space-times}\label{section:oneparameter}

Let us first discuss metrics which depend only on one dimension-full parameter, i.e. a single mass scale. This parameter defines a family of metrics and we are interested in measuring the distance between two representatives of such a family.

\vskip0.3cm
The discussion in Section \ref{section:generalExamples} can be immediately applied to this case. Indeed, for dimensional reasons, the metric can be put in the form
\be
g_{\mu\nu}(x,M)=M^{-2}\tilde g_{\mu\nu}(x) \ ,
\label{oneScaleSpaces}
\ee
by using dimensionless coordinates $x^\mu$. It then follows from (\ref{conformalAnsatz}-\ref{distanceConformal}) that
\be
\Delta_g = 2c\sqrt{d}\log(\frac{M_i}{M_f})\;.
\ee
As we already said, this result agrees with calculations made in \cite{Lust:2019zwm} regarding the AdS distance conjecture. We will now recover it for specific one parameter space-times.

\subsection{de Sitter space distance}

Let us first consider briefly the case of de Sitter space. The details are very close to the ones encountered in \cite{Lust:2019zwm}, {but we repeat them since this section will feature the first occurence of an infinite entropy limit as an infinite distance limit, preparing the ground for later discussions of black hole geometries}.

\vskip0.3cm
As it is well know, de Sitter space possesses a finite cosmological horizon and the $d$-dimensional de Sitter metric in static coordinates can be written as
\be
\text ds^2 = -\biggl(1-{r^2\over r_S^2}\biggr)\text dt^2+ \biggl(1-{r^2\over r_S^2}\biggr)^{-1}\text dr^2+r^2\text dS_{d-2}^2\;.
\label{static}
\ee
The horizon radius $r_S$ is related to the positive de Sitter cosmological constant $\Lambda_\text{dS}$ by
\be
\Lambda_\text{dS}={(d-1)(d-2)\over 2r_S^2}\;.
\ee
We then measure the distance between two de Sitter space-times of cosmological constants $\Lambda_i$ and $\Lambda_f$ by solving \eqref{geodesicsVolume}. We choose for instance a solution ansatz for which the initial metric $g(\lambda_i)$ in the flow reproduces \eqref{static} with $r_S=r_S(\lambda_i)$, and similarly for $g(\lambda_f)$. However with this line element the space-time integrations present in \eqref{geodesicsVolume} complicate the analysis. This comes about because this set of coordinates is singular at the horizon. Instead, we use global coordinates,
\be
\text ds^2 = -\frac{r_S^2}{\cos^2(\eta)}\left(-\text d\eta^2+\text d\psi^2+\sin^2(\psi)\text dS_{d-2}^2\right)\;.
\label{globaldS}
\ee
The computation of the distance for this kind of space-times has already been carried out in Section \ref{section:generalExamples}, with \eqref{globaldS} being an explicit realization of \eqref{conformalAnsatz}. In addition, since de Sitter spaces are fully determined by their cosmological constants, the discussion around \eqref{oneScaleSpaces} also applies. Thus, \eqref{globaldS} represents a realization of those previous considerations in terms of well-behaved coordinates for which the computation is well defined. Eventually, we obtain the de Sitter distance
\be
\Delta_\text{dS}\sim\log r_S\sim -\log\Lambda_\text{dS}\;,
\label{desitterdistance}
\ee
as already determined in \cite{Lust:2019zwm}. 

\vskip0.3cm
Alternatively for de Sitter space, being an Einstein space of
constant curvature
\be
R_\text{dS}={d(d-2)\over r_S^2}\;,
\ee
 one can also follow the Ricci-flow along the space of metrics $g_{\mu\nu}(\Lambda_\text{dS})$. Then it turns out that $r_S\rightarrow\infty$ corresponds to an infinite distance fixed point of the Ricci flow and 
 the Ricci distance (\ref{riccidistance})
 agrees with the distance given in (\ref{desitterdistance}).
 
\vskip0.3cm
 Let us recall that the Gibbons-Hawking entropy \cite{Gibbons:1977mu} is proportional to the area of the event horizon, namely
\begin{equation}
{\cal S}_\text{dS}=1/\Lambda_\text{dS} \,
\end{equation}
Therefore we can easily express the distance in terms of the de Sitter entropy as 
\be
\Delta_\text{dS}\sim\log{\cal S}_\text{dS}\;,\label{desitterentropdistance}
\ee
and the limit of infinite entropy is at infinite distance in the space of de Sitter geometries.\footnote{The infinite distance at vanishing cosmological constant for AdS spaces was argued in \cite{Lust:2019zwm} to announce the presence of Kaluza-Klein modes from compact dimensions, as clearly seen in string constructions. The existence of such constructions for dS being unclear, similar arguments are not available to explain the infinite distances encountered in this section. However, the latter being driven by infinite entropies, it would be interested, although beyond the scope of this paper, to generalize the arguments to be developed in what follows for black hole geometries to the case of dS spaces.} This conclusion also holds for black hole geometries, which are the main focus of this paper and to which we now turn.

\subsection{Schwarzschild black hole}\label{section:schBH}

We now consider asymptotically flat geometries with finite  horizon. These are nothing else than black hole geometries.
Let us first consider the well-known $4$-dimensional Schwarzschild black hole of mass $M$. The metric reads
\be
\text ds^2 = -\biggl(1-{r_S\over r}\biggr)\text dt^2+ \biggl(1-{r_S\over r}\biggr)^{-1} \text dr^2+r^2\text dS_{2}^2\;,
\label{staticbh}
\ee
with horizon size $r_S=2M$.
Since 
 the Schwarzschild metric is Ricci flat, we cannot use the Ricci flow distance in order to compute $\Delta_{bh}$.
As in the case of de Sitter space, the choice of coordinates \eqref{staticbh} is not the one for which the distance ought to be  computed. Instead, we can use Kruskal-Szekeres coordinates, in which the metric reads
\be
\text ds^2=-4\frac{r_S^3}{r}e^{-\frac{r}{r_S}}\text dT^2+4\frac{r_S^3}{r}e^{-\frac{r}{r_S}}\text dR^2+r^2\text dS_{2}^2 \ ,
\label{bhKScoos}
\ee
with $\left(1 - \frac{r}{r_S}\right) e^{\frac{r}{r_S}} = T^2 - R^2$. This again leads to 
\be
\Delta_{bh}\sim \log r_S\, .
\ee
The result again agrees with the analysis of \eqref{oneScaleSpaces}, as it should since Schwarzschild geometry is fully specified in terms of the black hole mass. Indeed, \eqref{bhKScoos} is a realization of \eqref{conformalAnsatz} once we redefined $r\rightarrow r_S\times r$, or equivalently $r/r_S$ is fully specified in terms of the dimensionless coordinates $T$ and $R$. As for the dS global coordinates, \eqref{bhKScoos} offers a realization of well-behaved coordinates for which the computation is well defined and agrees with our general considerations for one-parameter space-times.

\vskip0.3cm
Expressed in terms of the black hole entropy the result reads
\be
\Delta_{bh}\sim \log {\cal S}_{bh}\, .
\ee

Instead of the full Schwarzschild geometry we can also consider the near horizon geometry. For that
we  introduce the  coordinate $\epsilon=r-r_S$ and for small $\epsilon$ we obtain the metric in the near horizon limit 
  \begin{equation}
\text ds^2=-{\epsilon\over r_S}\text dt^2+{r_S\over\epsilon}\text d\epsilon^2+r_S^2\text dS_2^2\,. 
\label{nearHorizonMetricSchwarzschild}
 \end{equation}
The distance when varying the horizon size is again
\begin{equation}
 \Delta_{bh}\sim\log r_S\, .\label{bhdistance}
 \end{equation}
At large distances, meaning at large black hole horizon, the near-horizon metric approaches Minkowski space,\footnote{Defining new coordinates $\rho=\sqrt{r_S\epsilon}$ and $\omega=t/r_S$, \eqref{nearHorizonMetricSchwarzschild} is the metric of $M^{1,1} \times S^2$, where $M^{1,1}$ is the 2-dimensional Minkowski space in Rindler coordinates. It flows at large $r_S$ to the metric of 4D Minkowski space in Rindler coordinates.} whose entropy is infinite \cite{Dvali:2015rea,Averin:2016ybl}.

\section{Two-parameters space-times}\label{sec5}

We now turn to the study of space-times defined by two dimensionful parameters.

\subsection{A prescription to compute the distance}\label{section:prescription}

For two-parameters space-times the discussion of section \ref{section:generalExamples} does not straightforwardly apply. Indeed, for a metric $g$ defined by two mass parameters $M_1$ and $M_2$, dimensional-analysis arguments only teach us that the metric can be put in the following form
\be
g_{\mu\nu}(x,M_1,M_2)=M_1^{-2}\tilde g_{\mu\nu}\left(x,\frac{M_1}{M_2}\right) \ ,
\ee
for dimensionless coordinates $x$. Thus, the geodesic equation and the distance depend on the precise functional form of $g$.

\vskip0.3cm
Nonetheless, in this paper we only study $4$-dimensional static space-times with two Killing vectors, for which the metric can always be brought in the form
\beq
\text d s^2=-V(r,\alpha_i) \ \text dt^2+\frac{\text dr^2}{V(r,\alpha_i)}+r^2 \text dS_2^2 \ ,
\label{2Kstaticmetric}
\eeq
where $V$ is a given function and the $\alpha_i$ dependence indicates the possible presence of parameters defining the solution. For such space-times, we compute the distance as follows: first, we reach the Eddington-Finkelstein gauge performing the transformation
\beq
\text d t= \text dv-\frac{\text dr}{V(r,\alpha_i)} \ ,
\eeq
which brings the metric to the form
\beq
\text d s^2=-V(r,\alpha_i) \ \text dv^2+2\text dv\text dr+r^2 \text dS_2^2 \ .
\label{EFmetric1}
\eeq
One can then rescale the $r$ and $v$ coordinates, $r \to \alpha\tilde r, v \to \alpha\tilde v$, where $\alpha$ is a function of the parameters\footnote{For metrics defined in terms of at least one mass scale $M$, $M=\alpha$ can be used to make the coordinates dimensionless, as we discussed in section \ref{section:generalExamples}.} $\alpha_i$. The metric becomes
\beq
\text d s^2=-\alpha^2 V(\alpha \tilde r,\alpha_i) \ \text d\tilde v^2+2\alpha^2 \text d\tilde v\text d\tilde r+\alpha^2 \tilde r^2 \text dS_2^2
\label{EFmetric2}
\eeq
Using this ansatz with $\alpha=\alpha(\lambda)$, the distance can be easily computed and yields
\beq
\Delta_g=4 \ c \log(\frac{\alpha(\lambda_f)}{\alpha(\lambda_i)}).
\eeq
This holds for every metric of the form \eqref{2Kstaticmetric} and any $\alpha$ since $V(\alpha \tilde r, \alpha_i)$ does not contribute when using \eqref{dismetr}.
 Our prescription to compute the distance is thus to bring the metric in the form \eqref{EFmetric2}, then to use it to solve the geodesic equation \eqref{geodesicsVolume} and to compute the distance \eqref{dismetr}. Note that the geodesic equation for the ansatz \eqref{EFmetric2} is not trivial, as is shown later on for specific cases.

\vskip0.3cm
The reader may wonder why we need to define such a precise prescription. The answer lies in the fact that \eqref{dismetr} is not compatible with diffeomorphisms which vary along the geodesic flow, so that the initial frame from which the flow is computed has to be specified, as explained in greater details in appendix \ref{appdiffeos}. At this point a legitimate question about the universality of our results can be raised. To answer this question we notice however that in all the examples studied, if we change the coordinates system, the geometric distance is either not computable -- due to problems in the coordinates integration range -- or again logarithmic in the parameter. Therefore, although not a proof, this is a confirmation that our prescription is sensible.

\subsection{Reissner-Nordstr\"om black hole}

\subsubsection{Non-extremal case}

The first example of a two-parameter family of solutions is the non-extremal Reissner-Nordstr\"om metric, which describes a charged black hole. It can be written as in \eqref{2Kstaticmetric} with 
\begin{eqnarray}
V(r,\alpha_i)=\frac{(r-r^+_S)(r-r^-_{S})}{r^2} \ .\label{staticchargedbh}
\end{eqnarray}
This geometry possesses two horizons $r^\pm_S$. In terms of the black hole mass $M$ and charge $Q$, the horizons positions are given by 
\be
r_S^\pm=M\pm\sqrt{\delta}\ , \text{ with } \delta=M^2-Q^2.
\ee
To avoid the appearance of a naked singularity we must impose $|Q|\leq M$, such that $r^+_S$ and $r^-_S$ are the outer and inner horizon, respectively.

\vskip0.3cm
We now apply the prescription of section \ref{section:prescription}, i.e. we rescale the coordinates using a dimensionful parameter $\alpha$ of the metric, so that we obtain a distance which scales logarithmically with respect to $\alpha$. In the current setup, $\alpha$ can be built combining $M$ and $Q$ in an arbitrary fashion: $\alpha=\alpha(M,Q)$. Thus, the infinite distance in $\alpha$ corresponds to some geodesic path in $(M,Q)$ space, which we now study. It is easy to see that, for the metric \eqref{EFmetric2}, the right hand side of \eqref{geodesicsVolume} vanishes, so that we are left with
\be
\dot{f}=0
\ee
as a geodesic equation.
For instance, the solutions for the choice $\alpha=M$ are
\be
M(\lambda)=c_1e^{\frac{\lambda}{4}} \ , \quad Q(\lambda)= c_3 \sqrt{2 \lambda -c_2}e^{ \frac{2 \lambda -c_2}{8}}
\label{MasAlphaRN}
\ee
(remember that we specialized to $d=4$). Imposing $\abs{Q}\leq M$, we must choose $c_3=0$ (and $c_1>0$) and we actually describe the flow of Schwarzschild black holes, with infinite distance points which are either zero or infinite mass black holes, consistently with the results of section \ref{section:schBH}.
If $\alpha=Q$, the solutions are
\be
M(\lambda)=(c_2\lambda+c_3)e^{\frac{\lambda}{4}} \ , \quad Q(\lambda)= c_1e^{\frac{\lambda}{4}} \ ,
\label{QasAlphaRN}
\ee
and the only (asymptotic) bound is $c_3>\abs{c_1}$ if $c_2=0$ (and $\text{sgn}(c_2)= \text{sgn}(\lambda)$ always). This case allows to describe the infinite distance limits of a zero ($\lambda\rightarrow-\infty$) or infinite ($\lambda\rightarrow+\infty$) mass and charge black hole, either with $\frac{Q}{M}$ fixed ($c_2=0$) or going to zero ($c_2\neq0$).\footnote{$M(\lambda)=c_2\lambda+c_3 \ , \ Q(\lambda)= c_1$ is also a solution, which corresponds to asymptotically large $M$ for fixed $Q$, with a zero associated distance. This describes a null geodesic, for which our notion of proper time and distance does not apply. We disregard this possibility and similar ones in what follows.}

\vskip0.3cm
The most natural choice for $\alpha$ is provided by the use of the entropy of the RN black bole,
which 
is proportional to the area of the outer event horizon,
\begin{equation}
{\cal S}_{RN}=\pi {r_S^+}^2\, .
\end{equation}
Therefore, in order to express the distance in terms of the black hole entropy,  we 
 take $\alpha=r_S^+$ as the relevant parameter and 
 follow the prescription explained before. This leads to the distance with respect to  $r_S^+$, where the ratio $r_S^+/r_S^-$ is kept fixed,\footnote{The fact that the ratio $r_S^+/r_S^-$ is kept fixed is found by solving the geodesic equation, which leads to
\be
r_S^+=c_1e^{c_2\lambda} \ , \quad r_S^-=(c_3+c_4\lambda)e^{c_2\lambda} \ ,
\ee
for an affine parameter $\lambda$. Demanding $r_S^-<r_S^+$ imposes $c_4=0$ (and $c_1>c_3$), so that eventually $r_S^+/r_S^-=c_1/c_3$. $c_2$ is fixed when demanding that $\lambda$ is the proper time.}
\be
\Delta_{RN}\sim \log r_S^+\sim \log{\cal S}_{RN}\, .\label{rndis}
\ee
This behaviour is also reproduced by \eqref{MasAlphaRN} and \eqref{QasAlphaRN} up to possible logarithmic corrections. Indeed, the first correction scales as $\log \log S_{RN}$, if $\alpha=Q$ and $c_2\neq 0$.

\subsubsection{Extremal case}\label{RNEXT}
 
In the extremal limit ($Q^2=M^2$) we obtain
\begin{eqnarray}
V(r,\alpha_i)=\left(1-\frac{|Q|}{r}\right)^2
\end{eqnarray}
Here the horizon is at $r_S=|Q|$ and the entropy simply becomes
\be
{\cal S}_{RN}=\pi Q^2\, .\label{srn}
\ee
Since the extremal RN black hole is given in terms of a single scale $Q$, there is no possible subtlety in the prescription and we are back to the analysis from section \ref{section:oneparameter}. The distance is thus:
\be
\Delta_{RN}\sim \log Q\sim \log {\cal S}\, .\label{rndisextr}
\ee

\subsection{AdS Schwarzschild black hole}

The AdS Schwarzschild black hole has
\begin{eqnarray}
V(r,\alpha_i)=1-\frac{2M}{r}+\frac{r^2}{\ell_S^2}
\end{eqnarray}
Choosing $\alpha=M$ we obtain the distance
\beq
\Delta_g\sim \log M
\eeq
and 
\beqn
M(\lambda) &=&c_1 e^{\frac{\lambda}{4}}\\
l_S(\lambda) &=& \frac{c_3 e^{\frac{\lambda}{4}}}{\sqrt{c_2+2 \lambda}},
\eeqn
whereas if $\alpha=\ell_S$,
\beq
\Delta_g\sim \log \ell_S
\eeq
and 
\beqn
M(\lambda) &=& \left(c_2 \lambda+c_3\right)e^{\frac{\lambda}{4}} \\
l_S(\lambda) &=& c_1 e^{\frac{\lambda}{4}}.
\eeqn
Depending on the values of the integration constants, different scenarios are depicted. For instance, it is possible that $M\to 0$ faster than $\ell_S\to 0$.

\vskip0.3cm
We can again rewrite the distance in terms of the entropy
\be
S_{bh}=\pi r_h^2
\ee
with $r_h$ the horizon radius implicitly defined by
\be
M=\frac{r_h}{2}\left(1+\frac{r_h^2}{l_S^2}\right) \ .
\ee
This implies that asymptotically
\be
r_h\sim (Ml_S^2)^{\frac{1}{3}} \text{ and } \Delta_g\sim \log S_{bh}  \ .
\ee
Again, there might be logarithmic corrections to the distance written in terms of the entropy, for instance if $c_2\neq0$ when $\alpha=l_S$.

\section{${\cal N}=2$ extremal black holes, infinite distance and Kaluza-Klein modes as infinite tower}\label{section:n=2}

{Having established that the distance scales logarithmically with the black hole entropy, so that infinite distances are limits of infinite entropies, we turn to the study of candidates for the tower of states announced by the BHEDC in \eqref{BHEDCstates}.}

In this section we consider ${\cal N}=2$ extremal black holes in string theory.
In the large entropy limit some of the  black hole  charges  must become very large.
 Via the ${\cal N}=2$ attractor mechanism \cite{Ferrara:1995ih,Ferrara:1996dd}, {one could expect} that some of the internal moduli fields also have to take large field values and that the 
associated Kaluza-Klein modes become very light,
{thus providing the states in the BHEDC tower. However, we will see that this intuition is not always realized.}

\vskip0.3cm
So let us consider four-dimensional extremal, ${\cal N}=2$ supersymmetric black holes in string theory.
E.g. in type IIA string compactifications, four-dimensional, extremal black holes with non-vanishing entropy can be constructed by the intersection of $N_V$ D4-branes, which are wrapped around four-cycles 
of the internal six-dimensional Calabi-Yau manifold. In addition to the D4-branes one also needs a certain number of D0-branes.
The four-dimensional effective field theory is given in terms of ${\cal N}=2$ supergravity theory coupled to $N_V$ ${\cal N}=2$ vector multiplets.
The corresponding Abelian gauge group is given by
\begin{equation}
G=U(1)^{1+N_V}\, ,
\end{equation}
where the additional $U(1)$ factor comes from the graviphoton gauge symmetry.
The four-dimensional metric of   ${\cal N}=2$  extremal black hole solutions has the following form:
 \be
\text ds^2 = -\biggl(1-{\sqrt{A/4\pi}\over r}\biggr)^2\text dt^2+ \biggl(1-{\sqrt{A/4\pi}\over r}\biggr)^{-2}\text dr^2+r^2\text dS_{2}^2\;,\label{N=2bh}
\ee
where $A$ is the area of the event horizon.
Via the so-called attractor mechanism \cite{Ferrara:1995ih,Ferrara:1996dd}
the corresponding entropy is determined by the extremization of the central charge $Z$ of the ${\cal N}=2$ supersymmetry algebra on the horizon. 
Specifically,  the entropy takes the following form \cite{Behrndt:1996jn}:
\begin{equation}
{{\cal S}_{{\cal N}=2}\over \pi}={A\over 4\pi}=|Z_{\rm hor}|^2=\sqrt{-4Q_0d_{ijk}P^iP^jP^k}\,.\label{n=2entr}
\end{equation}
Here $Q_0$ is the electric charge of the black hole with respect to the graviphoton $U(1)$ gauge group, and the $P^i$ are magnetic charges with respect to $U(1)^{N_V}$.
Moreover $Q_0$ corresponds to the number of D0-branes, and the magnetic charges $P^i$ are the wrapping numbers of the D4-branes around the corresponding internal four-cycles.
Finally $d_{ijk}$ denotes the triple intersection numbers of the four-cycles.
The entropy formula \eqref{n=2entr} is the ${\cal N}=2$ string generalization of the Reissner-Nordstr\"om entropy  given
in \eqref{srn}.  ${\cal S}_{{\cal N}=2}$ becomes large in the limit of large electric charge $Q_0$ or 
large magnetic charges $P_i$,
and, as we have shown in section \ref{RNEXT}, the limit of large charges is at infinite distance in the space of extremal ${\cal N}=2$ metrics. 

\vskip0.3cm
Let us consider the solutions of the attractor equations, which tell that the value of the scalar fields  at the black hole horizon are entirely determined by the electric and magnetic charges in the following way.
For the K\"ahler moduli 
$T_i$,
 which determine the sizes of the two-cycles perpendicular to the wrapped four-cycles of the D4-branes, one obtains
\begin{equation}
T^i=P^i\sqrt{-{Q_0\over d_{jkl}P^j P^kP^l}}\, .
\end{equation}
In addition the ten-dimensional IIA dilaton $e^{-2\phi^{(10)}}$ is determined by the attractor equations as 
\begin{equation}
e^{-2\phi^{(10)}}=\sqrt{-{d_{ijk}P^i P^jP^k\over Q_0^3}}\, .
\end{equation}
In terms of  the dilaton, the black hole entropy can be also written as
\begin{equation}
{{\cal S}_{{\cal N}=2}}=2\pi Q_0^2e^{-2\phi^{(10)}}\,.\label{n=2entr1}
\end{equation}
The volume ${\cal V}$ of the internal
space is given as
\begin{equation}
{\cal V}=\,
d_{ijk} T^i T^j T^k=
\sqrt{-{Q_0^3\over d_{ijk}P^i P^jP^k}}
\, .
\end{equation}
The four-dimensional dilaton,
\begin{equation}
e^{-2\phi^{(4)}}=e^{-2\phi^{(10)}}{\cal V}=e^{-2\phi^{(10)}}
d_{ijk} T^i T^j T^k\, ,
\end{equation}
 independent of the charges $Q_0$ and $P^i$. 
In this case the four-dimensional Einstein frame is identical to the string frame, and the masses of the string states in IIA are also independent of the charges.\footnote{In a dual heterotic
compactification on $K3\times T^2$ one of the IIA moduli $T_i$ corresponds to the heterotic dilaton field. Then the limit of large charges and large entropy corresponds
to a very weakly coupled string, and the tower of states are given by light tensionless strings.}
Specifically for the Kaluza-Klein mass scale one obtains:
\begin{eqnarray}
m_\text{KK}&=&{M_s\over {\cal V}^{1/6}}={M_p\over 
(d_{ijk} T^i T^j T^k)^{1/6}}\nonumber\\&=&
{ {(d_{ijk}P^iP^jP^k})^{1/12}\over (-Q_0)^{1/4}}
\, .\label{KKscale}
\end{eqnarray}

\vskip0.3cm
Let us now compare the Kaluza-Klein mass scale \eqref{KKscale} with the entropy \eqref{n=2entr} of the extremal ${\cal N}=2$ black hole:
\begin{equation}
m_\text{KK}=\biggl({1\over {\cal S}_{{\cal N}=2}}\biggr)^{1/2}     2{(d_{ijk}P^iP^jP^k})^{1/3}  \, .\label{attractor}
\end{equation}
We see that,
{since the magnetic charges are quantized,}\footnote{For small magnetic charges $P^i$, the higher curvature corrections to the
black hole entropy, which are proportional to the second Chern class $c_2$ of the Calabi-Yau manifold, are small.} the 
black hole entropy becomes very large when the Kaluza-Klein masses are light.
Specifically for fixed  $P^i$, the mass scale of the tower of states is given in terms of a square root of the inverse entropy, i.e.~one gets that $\alpha=1/2$ for the power in 
\eqref{gendisc}.
This relation between the relevant mass scale and the entropy is identical to the relation between the tower mass scale and the cosmological constant $\Lambda$ in the strong 
AdS Distance Conjecture  (ADC)   \cite{Lust:2019zwm}. 
On the other hand, 
there are also choices of electric and magnetic charges, for which   $m_\text{KK}$  can be kept constant, but the entropy becomes very large.

{So the relation between $m_\text{KK}\rightarrow0$ and ${\cal S}_{{\cal N}=2}\to\infty$ depends in what (infinite) direction one goes in the  $(Q_0,P^i)$ charge-space.
For simplicity let us consider the case of three magnetic charges and set them equal to each other, $P^i=P$. Furthermore it is convenient to replace $-Q_0$ by $Q$.
For the two-derivative action we need that $Q>P\geq 1$.
Then one obtains for the  comparison between the  Kaluza-Klein mass scale and the entropy the following expression:
\begin{equation}
m_\text{KK}=\biggl({1\over {\cal S}_{{\cal N}=2}}\biggr)^{1/2}   2P  \, .\label{attractor11}
\end{equation}
We see that $m_\text{KK}\rightarrow0$ implies that ${\cal S}_{{\cal N}=2}\rightarrow \infty$, since $P$ cannot be smaller than one.
So infinite distance in the internal moduli space always implies large entropy.
However, as said before, the converse is not true:
the limit ${\cal S}_{{\cal N}=2}\rightarrow \infty$ can also be realized for finite, non-vanishing $m_\text{KK}$, namely for $Q, P\rightarrow\infty$, but keeping the ratio $Q/P$ fixed.}

\vskip0.3cm
In summary, denoting the Kaluza-Klein mass scale as $m_\text{KK}=1/R$, we have seen that
\begin{equation}
\Delta(R)\sim\log R\rightarrow\infty\,\,\Longrightarrow\,\, \Delta({\cal S})\sim\log {\cal S}\rightarrow\infty\, . \label{KKtoS}
\end{equation}
The emergence of an infinite tower of massless states in the infinite distance limit of large internal moduli is consistent with the swampland idea.
Recall that the moduli fields can have arbitrary values at spatial infinity, but their horizon values must be expressed by the electric and magnetic black hole charges due to the attractor mechanism.
Therefore in the limit of large internal moduli, the effective horizon field theory must contain an infinite tower of states, namely the light tower of internal Kaluza-Klein states.
Without these light states, the effective field theory on the black hole horizon would be inconsistent, i.e. it would be lying in the swampland.
In fact, the near horizon geometries of extremal ${\cal N}=2$  extremal black holes are given by the space
AdS$_2\times S^2$ with so-called Bertotti-Robinson metric.  In the large moduli  limit, the effective, four-dimensional AdS$_2\times S^2$ theory must contain an infinite tower of states, which are just the 
light Kaluza-Klein modes from the internal Calabi-Yau manifold. {The reverse statement to \eqref{KKtoS} does not hold however,
\begin{equation}
\Delta({\cal S})\sim\log {\cal S}\rightarrow\infty\,\,\slashed{\Longrightarrow}\,\, \Delta(R)\sim\log R\rightarrow\infty \, ,
\end{equation}
and we must look for an other candidate for the BHEDC tower when the moduli are fixed but the entropy grows.}

\vskip0.3cm
{This conclusion regarding the status of Kaluza-Klein modes with respect to the BHEDC can also be extended to Calabi-Yau compactifications of string theory to four dimensions without any branes and hence without any electric and magnetic charges. The corresponding effective four-dimensional supergravity theory admits Schwarzschild solutions for every value of the moduli, hence Schwarzschild x CY is a perfectly fine (non-supersymmetric) string theory background. In particular, we can choose values of the moduli such that there are no light Kaluza-Klein modes, then take the limit $S\to \infty$ while keeping the moduli constant.}


\section{Non-extremal black holes and microstates as infinite tower}\label{sec6}

In previous sections we have seen that for Schwarzschild and for non-extremal Reissner-Nordstr\"om black holes
the
limit of 
infinite horizon $r_S\rightarrow\infty$, or equivalently the limit of infinite entropy ${\cal S}\rightarrow\infty$,
is at infinite distance in 
 the space of black hole metrics. 
 Via the infinite distance conjecture we now would get a prediction, namely the BHEDC stating that there is an infinite tower of modes, whose masses scale with a certain power of entropy as 
 \be
m_{\cal S} \sim {\cal S}^{-\alpha}\;,
\label{gendisc1}
\ee
where $\alpha$ is a positive constant. 
For large horizons and correspondingly for large entropies this mass scale becomes very tiny and goes to zero in the infinite entropy limit.  This is the same result we have found for extremal black holes. However, while for the latter one could rely on higher dimensional string-theoretical constructions, for non-extremal black holes we are bound to an intrinsic analysis in four dimensions in terms of properties of the black hole.

\vskip0.3cm
In this section we thus 
try to provide an interpretation of the infinite tower of modes  in terms of black hole horizon pseudo Goldstone  modes, which are closely related to the microstates of the black hole.
Concretely, besides the black hole microstates, all with masses of the order $M$, there are additional light modes that describe the transition between the different black hole microstates, i.e.
the excitations from one microstate into another one. They are these light modes, which we like to identify with the light tower of modes from the infinite distance conjecture.
{It is of course known that any system in the limit of infinite entropy must possess microstates, whose number goes to infinity together with the entropy. However a rather non-trivial question is what happens to the mass differences (frequencies) between these microstates? Here the infinite distance limit for the horizon suggests  that those frequencies, associated to modes of the theory, must vanish with a certain power of the entropy. That is, the  energy gap between each microstate must scale as ${\cal S}^{-\alpha}$.}
 In the following we like to argue that this behaviour is true when considering the Goldstone modes of black holes in terms of soft hair and related BMS-like
 diffeomorphisms on the horizon of a black hole.
 In other words, we  claim that the infinite tower of massless modes, which follows from the infinite horizon distance conjecture, corresponds to the Goldstone modes 
 of the BMS-like transformations on the black hole horizon. We stress again that these modes connect a microstate to another. If the microstates are completely degenerate and infinite, then the energy of a mode that brings from  a state to another is zero, and there are infinitely many modes. {Our claim follows from results on BMS transformations in classical gravity, as well as their interpretation in a quantum model of black holes, the so-called N-portrait. We review those two aspects below, and eventually link them to the BHEDC.}
  
 \vskip0.3cm
 As by now advocated  in many papers     
  \cite{Donnay:2015abr,Blau:2015nee,Donnay:2016ejv,Lust:2017gez,Averin:2018owq,Donnay:2019zif,Donnay:2019jiz,Averin:2019zsi},
in the presence of an event horizon, the classical symmetry algebra of a black hole geometry is enhanced, 
 and it is this enhancement that is responsible for microstates of the black hole. 
 So let us recall the classical BMS-like transformations on the black hole horizon. For this it is useful to choose 
  Eddington-Finkelstein coordinates, so that the Schwarzschild metric gets the form
\begin{equation}\label{Sch}
\text ds^2 = -\left(1-\frac{r_S}{r}\right)\text dv^2 + 2\text dv\text dr + r^2 \text d\Omega^2.
\end{equation} 
 Performing the change of variable
\be
r=r_S(1+2 \kappa \rho),
\label{map}
\ee
where $r_S=1/(2\kappa)$ for Schwarzschild, and expanding in powers of $\rho$, one finds that \eqref{Sch} becomes
\begin{equation}
\text ds^2=-2 \kappa \rho \ \text dv^2+2\text d\rho\text dv +\left(\frac{1}{4\kappa^2}+\frac{\rho}{\kappa}\right)(\text d\theta^2+\sin^2 \theta\text d\phi^2) \ ,
\label{change}
\end{equation}
up to $O(\rho^2)$ terms. This metric belongs to a larger class of near horizon metrics. 
Using  this metric one can compute 
the asymptotic Killing vectors $\zeta_f$, i.e. the set of vectors preserving the falloffs of the metric on the horizon. Using \eqref{map} one can then rewrite this result in terms of $r$ and $r_S$ to get
\be
\zeta^\mu_f=\left(f(\theta,\phi),0,\frac{\partial f}{\partial \theta}\left(\frac{1}{r}-\frac{1}{r_S}\right),\frac{1}{\sin^2 \theta}\frac{\partial f}{\partial \phi}\left(\frac{1}{r}-\frac{1}{r_S}\right)\right)\, ,
\label{vfL}
\ee
where $f(\theta,\phi)$ is an arbitrary function on the two-sphere.
As shown in \cite{Averin:2016ybl,Averin:2016hhm}, these are precisely the diffeomorphisms acting on the Schwarzschild metric that leave the horizon and the black hole mass $M$ invariant. 
The algebra of the asymptotic Killing vectors of the Schwarzschild black hole becomes an infinite-dimensional, commutative supertranslation BMS algebra.\footnote{One can derive another set of supertranslations and superrotations but we focus only on \eqref{vfL} here.}
 
 \vskip0.3cm
 In the classical limit $\hbar \rightarrow 0$ the entropy and hence number of black hole microstates becomes
 infinite. This may sound a bit paradoxical, but it means that the classical ground state of a black hole is infinitely degenerate.
 All the infinite number of classical microstate geometries have mass $M$, and they are completely degenerate.  They are all
 connected to each other by the BMS-like transformations  (\ref{vfL}) on the black hole horizon. 
 Choosing one of the infinitely many classical
 black hole ``ground states" with mass $M$, the event horizon symmetries are spontaneously broken. The associated Goldstone modes are massless
 and describe the transitions among the infinitely many microstate geometries. In fact one can compare the different vacua of the black hole geometry with the infinitely many Higgs vacua
for the case of spontaneous symmetry breaking. Each value of the Higgs field corresponds to a black hole microstate of mass $M$, and the classical, massless Goldstone
modes correspond to the transitions from one Higgs vacuum to another one. 

\vskip0.3cm
As we will mention at the end, there are also Higgs systems with non-vanishing entropy, where the
distance  to some critical Higgs field value in the scalar field space and the associated entropy  become infinite.
The black hole entropy becomes infinite in the semiclassical limit of large horizon, i.e. $r_S \to \infty$.
Performing this    takes the future horizon at $r=r_S$ to past null infinity ${\rd I}^-$.  
In this limit the vector fields (\ref{vfL}) reduce to
\be
\zeta^\mu_f=\left(f(\theta,\phi),0,\frac{\partial f}{\partial \theta}\frac{1}{r},\frac{1}{\sin^2 \theta}\frac{\partial f}{\partial \phi}\frac{1}{r}\right),
\ee
which are the standard BMS supertranslation vector fields at null infinity. In this limit the associated Goldstones modes at the horizon coincide with the standard BMS modes, which are massless
and are closely related to the soft graviton modes at ${\rd I}^-$.
So for infinite horizon, the massless tower of states from the infinite distance conjecture should coincide with the infinitely many true BMS modes at ${\rd I}^-$.

\vskip0.3cm
Now we like to argue that in the quantum case the different black hole microstates and their associated entropy carriers exhibit the right tower behaviour of the infinite distance conjecture.\footnote{Related arguments for the case of de Sitter potential via towers of particles and species as entropy carriers were provided in \cite{Ooguri:2018wrx}.} {For that we need first to provide a quantum picture of a black hole, which we take to be the N-portrait model of \cite{portrait,portrait1,Dvali:2013eja,portrait3,portrait4,gold}. 
There (some more details about black hole microstates, the N-portrait models and Goldstone modes are provided in appendix \ref{appC}), a black hole is understood as a self-sustained bound state of $N$ gravitons, where $N$ is the black hole entropy. The model provides several corpuscular interpretations, e.g. of Hawking radiation, but the most important for us is that the black hole is understood as a critical system due to the self-sustained character of the bound state and the energy dependence of the gravitational effective coupling. Then, by analogy to the physics of critical Bose-Einstein condensates in condensed matter, one concludes that there exist light collective excitations (dubbed Bogoliubov modes) of the bound gravitons composing the black hole. Furthermore, these collective modes can be identified in the classical limit to BMS modes \cite{Dvali:2015rea,Averin:2016ybl,Averin:2016hhm}.} In the quantum theory, namely in case of finite $\hbar$ and finite $r_S$, the BMS-like symmetries on the horizon are explicitly broken, and the (Bogoliubov-)Goldstone modes acquire masses, i.e.
they become pseudo Goldstone modes. In the N-portrait model, this means that at the quantum level the black hole microstates are not any more completely degenerate, but they possess masses of the order $M+m_l$, where the mass differences $m_l$
agree with the masses of the pseudo Goldstone bosons and go to zero in the limit of large $r_S$ or $\hbar\rightarrow 0$. Hence the transition among the non-degenerate black hole microstates are 
given in terms of the  pseudo Goldstone excitation modes with finite masses   $m_l$. It therefore costs a finite amount of energy to go from one black hole quantum microstate to another one.

\vskip0.3cm 
{The presence and behaviour of the pseudo Goldstone modes when the entropy grows resonate nicely with the conclusions of the BHEDC. So let us now discuss the precise scaling of the masses of those modes, namely let us ask} what is now the quantum spectrum of these pseudo Goldstone modes?
 Without a microscopic quantum theory that could tells us what is the energy difference of going from one microstate to another one,
 one can only make some guesses. For example, the modes in the Hawking radiation can be viewed as entropy carriers. These have typical frequencies 
 of the order of the Hawking temperature $T_H\sim 1/M$, and one could conclude that
 the mass gap is of order
   \be
m \sim {\hbar\over r_S}\sim\biggl({1\over {\cal S}}\biggr)^{1/ 2} \, .
\ee
However, based on the role of quantum criticality and entropy scaling of the N-portrait model,
 there must be an additional entropy suppression by a factor $1/{\cal S}$, which takes into account the relative deviations
from the thermal spectrum in the Hawking radiation.
It follows
that the mass gap is bounded to be at most
  \be
m_{\rm max} \sim {1\over {\cal S}}{\hbar\over r_S}  \sim\biggl({1\over {\cal S}}\biggr)^{3/ 2}    \, .\label{maxm}
\ee
Therefore in terms of the black hole mass $M$, the energy gap has to satisfy $m_{\rm max}\leq 1/M^{3}$.
Finally, in order to get the right number of microstates, 
  it was argued in    \cite{Averin:2016ybl} (see also \cite{Wen:2019bjp}) that
 the masses of the pseudo Goldstone modes have to obey the following spectrum:
 \be
m_l \sim {l^2\over {\cal S}^2}{\hbar\over r_S}\sim {l^2m_0}\;, \ {\rm with} \ l=1,\dots ,l_{\rm max} \ {\rm and} \  l_{\rm max}=\sqrt{\cal S}\, .
\ee
The quantum number $l$ can be viewed as an angular momentum number, and hence each mode has an $l^2$ degeneracy. 
Thus, the total number of modes contributing to the entropy scales as $l_{\rm max}^2={\cal S}$. 
Furthermore the maximal energy gap is indeed given by \eqref{maxm}.
From these formulas we can easily read off the associated microscopic mass scale, namely the mass of the lowest pseudo Goldstone excitation expressed in Planck units:
 \be
m_0 \sim \biggl({1\over {\cal S}}\biggr)^{5/ 2}\;.
\label{gendisc2}
\ee
 So we see that the tower of the infinite distance conjecture for the entropy in \eqref{gendisc}
 and the microscopic tower of the pseudo Goldstone bosons agree for the value $\alpha=5/2$. 
  In the context of the swampland scenario, the emergence of  the light pseudo Goldstone modes 
  in the infinite entropy limit of non-extremal black holes means that the
 effective horizon field theory must contain these states. In other words,
without these light states, the effective field theory on the black hole horizon would be inconsistent, i.e. it would be lying in the swampland.
Notice that the argument depicted here requires only the fact that quantumly the microstates are finitely many and it costs energy to move from one microstate to another. This property does not necessarily demand the microstates degeneracy to be lifted. Thus our argument is valid also in other models, like e.g. the fuzzball proposal (see the reviews \cite{Mathur:2005zp, Bena:2007kg} and references therein), although the details of the latter differ from the N-portrait model.

\vskip0.3cm
Let us also comment on what is  happening with the pseudo Goldstone modes in the extremal limit of a charged RN black hole.
Towards the extremal case with $T_H=0$ and with ${\cal S},M=|Q|\rightarrow\infty$ we expect that the Goldstone modes
become massless, although the entropy of the extremal black hole is still finite. Indeed, the mass gap of the Goldstone modes is zero for extremal black holes even in the quantum regime, because their masses are proportional to the black hole temperature.
{The nature of the tower of states for extremal black holes, and its relation with the one of non-extremal black holes, is still an open question and source of investigation. In section \ref{section:n=2}, we failed to identify it in terms of internal light Kaluza-Klein modes or light string excitations from a higher dimensional construction.}


\section{Conclusion}\label{sec7}

We have argued that besides the physical Kaluza-Klein modes and string excitations there exists another kind of tower of modes that can be determined
by the  distance  to the large entropy limit in the space of non-extremal black hole geometries. This tower of modes is closely related to the black hole microstates.
It is given in terms of pseudo Goldstone modes, which lift the degeneracy of the black hole microstates in an N-portrait model and describe the transitions among them.
These modes are part of the lower-dimensional effective field theory on the horizon of the black hole.
According to the swampland distance conjecture, this tower of states is necessary to have a consistent effective theory in quantum gravity on the black hole horizon. 
In other words, without this tower of states, the black hole with large entropy would belong to the swampland.
In fact, we believe that the entropy distance (\ref{entropydistance}) and the associated tower behaviour (\ref{gendisc}) are  general features of black holes in quantum gravity and their microstates.
Applied to macroscopic black holes like supermassive Kerr  black holes  with a horizon scale that is much larger than the Planck length $L_p$, the associated tower of states on the black hole horizon is enormous, reflecting
the fact that the entropy of macroscopic black holes is also huge.
However, since these modes  live on the horizon,  they are hard to access by earth experiments.

\vskip0.3cm
{For extremal, BPS black holes we have seen that the large entropy limit at infinite distance cannot always be related to a tower of states, which are given in terms
  of physical Kaluza-Klein modes of an internal Calabi-Yau manifold. The nature of the BHEDC tower for extremal black holes is thus left as an open question. Nonetheless, we can still say that large entropies follow from small Kaluza-Klein masses by the underlying attractor equations of ${\cal N}=2$ supersymmetric black hole solutions. It would be therefore interesting to see if there is a possible relation between attractor equations and the geometric flow equations.}


\vskip0.3cm
As  advocated in   \cite{Dvali:2019jjw} the behaviour of the ``mass" of entropy carriers compared to the entropy in {\eqref{maxm} is indeed 
universal and follows in any system with entropy from unitarity and also from the species bound. 
The invariant statement is the connection between the gap and the entropy. From the unitarity arguments,  the maximum energy gap is of the order of $ 1/{\cal S}$ (for black holes in units of the horizon), but it  could also be smaller. 
The distance in space of field theories with entropies is then given in terms of a field space  metric, which is proportional to $\log{\cal S}$.
For instance, in a Higgs field theory described in \cite{Dvali:2019mhn}, 
 the infinite entropy limit is at infinite distance in the field space.  This limit is reached for a critical value of the Higgs field, where the system reaches a quantum critical point.
  
\vskip0.5cm
\vspace{10px}
{\bf Acknowledgements}
\vskip0.1cm
\noindent

We thank Alex Kehagias and Marios Petropoulos for their collaboration at early stages of the project and Francesco Alessio, St\'ephane Detournay, Gia Dvali, Alavaro Herraez,
Eran Palti, Cumrun Vafa and Nick Warner
for useful discussions. We also thank the Corfu Summer Institute 2019, during which this project was initiated.
Q.B. is supported by the Deutsche Forschungsgemeinschaft under Germany's Excellence Strategy  EXC 2121 ``Quantum Universe" - 390833306.
The work of L.C. is supported by the ERC Advanced Grant ``High-Spin-Grav".
The work of D.L. is supported  by the Origins Excellence Cluster.
The work of S. L. is supported by the ANR grant Black-dS-String (ANR-16-CE31-0004).

\begin{appendices}
\section{The metric distance and diffeomorphism invariance}\label{appdiffeos}

The metric distance \eqref{dismetr} seems invariant under coordinate changes $x^\mu\rightarrow x^\mu(\tilde x)$ of the manifold $M$, since one integrates space-time scalars with the $\sqrt{g}$ measure. On the other hand, there is a $\partial_\tau$ derivative which may contradict the previous statement if one tries to change the coordinates in a $\tau$-dependent fashion. In this appendix we comment on these diffeomorphism-related issues.

\subsection{An Eddington-Finkelstein puzzle}\label{appEF}

We can easily understand the need to clarify the role of diffeomorphisms by looking at (4D) space-times in Eddington-Finkelstein (EF) gauge, discussed in section \ref{section:prescription}. We computed there the distance for the metric \eqref{EFmetric2}, and obtained
\beq
\Delta_g=4 \ c \log(\frac{\alpha(\tau_f)}{\alpha(\tau_i)}) \ .
\label{nonzeroDistance}
\eeq
However, we could have also computed it for \eqref{EFmetric1}, in which case we would have obtained
\beq
\Delta_g=0 \ .
\label{zeroDistance}
\eeq
One thus immediately sees that the distance depends on the coordinate choice, since those two results were reached for two metrics related by the change of coordinates $r \to \alpha(\tau)\tilde r, v \to \alpha(\tau)\tilde v$. 
{Similar puzzles would arise if one tried to identify the distance for a metric parameter from the kinetic terms of its fluctuations in different coordinate systems.}

\subsection{Fixed charts VS evolving charts}\label{appCharts}

The change between \eqref{nonzeroDistance} and \eqref{zeroDistance} can be mapped to the $\tau$ dependence of the distance $\Delta_g$. Indeed, by changing the EF coordinates according to $(r,v) \to \alpha(\tau)(r,v)$, we performed a diffeomorphism which changes along the metric flow. The distance is not invariant under such a diffeomorphism: let us consider a change of coordinates
\begin{equation}\label{eq:diffeo}
x^\mu =  x^{\mu}(\tilde x^\nu, \tau)=\tilde x^\mu+\xi^{\mu}(\tilde x^\nu, \tau) \,,
\end{equation}
so that the metric transforms according to
\begin{equation}
\tilde g(\tilde x^\mu, \tau) = \Phi^T(\tilde x^\mu, \tau) \, g(x^{\nu}(\tilde x^\mu,\tau), \tau) \,\Phi(\tilde x^\mu, \tau) \,,
\end{equation}
with
\begin{equation}
\Phi^\mu{}_\nu = \frac{\partial x^\mu}{\partial \tilde x^\nu}= \delta^\mu{}_\nu+\frac{\partial \xi^\mu}{\partial \tilde x^\nu} \,.
\end{equation}
Consequently, \eqref{dismetr} transforms non trivially since we have
\begin{equation}
\bal
\partial_\tau{\tilde g} =& \ \Phi^T \left(\partial_\tau g + \partial_\mu g \, \partial_\tau x^\mu \right) \Phi + \partial_\tau \Phi^T g \Phi + \Phi^T g \partial_\tau \Phi \\
=& \ \Phi^T \left(\partial_\tau g + {\cal L}_{\partial_\tau\xi} g\right)\Phi \,,
\eal
\end{equation}
where $\partial_\tau\xi^\mu$ is calculated at fixed $\tilde x$ and ${\cal L}_X$ is the Lie derivative with respect to the vector $X$ in coordinates $x$. We thus expect to map
\be
\Delta_{\tilde g}= c \int_{\tau_i}^{\tau_f}  \left( \frac{1}{V_M} \int\text d^4\tilde x \sqrt{\tilde g} \tr[\left(\tilde g^{-1}\frac{\partial \tilde g}{\partial \tau}\right)^2] \right)^{\frac12}\;,
\ee
to 
\be
\Delta'_g= c \int_{\tau_i}^{\tau_f}  \left( \frac{1}{V_M} \int \text d^4x \sqrt{g} \tr[\left(g^{-1}\left[\frac{\partial g}{\partial \tau}+{\cal L}_{\partial_\tau\xi} g\right]\right)^2] \right)^{\frac12}\;,
\ee
with $\xi$ defined in \eqref{eq:diffeo}.

We can test this formula on the EF coordinates of section \ref{appEF}: there, we had
\be
\xi^\mu=\big(\xi^{\tilde v},\xi^{\tilde r},\xi^{\tilde \theta},\xi^{\tilde\phi}\big)=\big((\alpha-1)\tilde v, (\alpha-1)\tilde r,0,0\big) \, ,
\ee
hence
\be
\partial_\tau\xi^\mu=(\partial_\tau\alpha \ \tilde v,\partial_\tau\alpha \ \tilde r, 0,0)=\left(\frac{\partial_\tau\alpha}{\alpha} \ v,\frac{\partial_\tau\alpha}{\alpha} \ r, 0,0\right)
\ee
and we get as expected $\Delta_{\tilde g}=\Delta'_g=4 c \log(\frac{\alpha(\tau_f)}{\alpha(\tau_i)})$.

We can also check from this formula that the distance is unaffected by diffeomorphisms which are constant along the flow, since then $\partial_\tau\xi=0$. A good example of this is the distance for the Schwarzschild black hole, which we know how to compute in Kruskal coordinates, as done in section \ref{section:schBH}, and in dimensionless EF coordinates as discussed in section \ref{section:prescription}. The two frames being related by M-independent diffeomorphisms, the computation of the distance leads to identical results in those two frames.

\subsection{The need for a preferred frame}

The previous discussion shows that before computing any distance, one should choose the reference frame in which the Lie derivative is absent from the distance. In section \ref{section:oneparameter} for instance, we chose to compute the distance with respect to dimensionless coordinates. This can be understood by saying that we chose coordinates which can be transformed to compact and geodesically complete ones without ever referring to the mass scale of the problem, i.e.~we used coordinates which correspond to charts which exist for any black hole/(A)dS space-times, irrespective of the mass/cosmological constant. This procedure is straightforward when only one mass scale is present.

\vskip0.3cm
In the case of a geometry defined by two dimensionful parameters, this procedure cannot be carried out since ratios of mass scales are dimensionless, so that there is an apparent degeneracy in choosing which combination of scales we should use when defining dimensionless coordinates. The identification of the relevant scale (assuming this exists) is also sometimes complicated by the need to use at least two charts to cover the maximally extended space-time, which is the case of the non-extremal RN black hole. Our strategy, defined in section \ref{section:prescription}, was then to try different choices for the mass parameter used to make the coordinates dimensionless, and identify a generic behaviour of the large distances in terms of large entropies.

\subsection{An alternative distance definition}\label{covdisapp}

Instead of using 
\be
\bal
\Delta_g = c \int_{\tau_i}^{\tau_f}  \left( \frac{1}{V_M} \int_M \sqrt{g} \tr[\left(g^{-1}\frac{\pa g}{\pa \tau}\right)^2] \right)^{\frac12}\text d\tau\;,
\label{dismetr1}
\eal
\ee
 following 
the conjecture B1 in \cite{Kehagias:2019akr}, one way to define the distance in the space of metrics $g$ is via the associated scalar curvature $R(g)$.
Specifically here the distance is defined as 
\begin{equation}
\Delta_g\simeq \log \Bigg({R(g^i)\over R(g^f)}\Biggr)\,.\label{resultR}
\end{equation}
Here $R(g^i)$ and $R(g^f)$ are the corresponding initial and final values of the scalar curvature.

At first sight, this definition for $\Delta_g$ appears to be useless for flat manifolds with $R(g)=0$, like the Schwarzschild black hole.
Actually for Ricci-flat spaces, one could try to introduce a regulator $\epsilon$ in the metric, which makes it non-flat, and at the end take the limit $\epsilon\rightarrow 0$. 
E.g. write the Schwarzschild metric in the following form
\be
\text ds^2 = -\biggl(1-{r_S\over r}\biggr)\text dt^2+ \biggl(1-{r_S\over r}\biggr)^{-1} \text dr^2+r^{2+\epsilon}r_S^{-\epsilon}\text dS_{2}^2\;.
\label{staticbh1}
\ee
The corresponding 
scalar curvature is
\begin{equation}
R(\epsilon) = \frac{4\left(\tfrac{r_S}{r}\right)^\epsilon r - r(2+\epsilon) (2+3 \epsilon) + r_S \epsilon (4 + 3 \epsilon)}{2 r^3}\, .\label{scalarc1}
\end{equation}
After changing coordinates $r=r_S\tilde r$, the above metric becomes
\be
\text ds^2 =r_S^2 \Biggl(-\biggl(1-{1\over \tilde r}\biggr)\text dt^2+ \biggl(1-{1\over \tilde r}\biggr)^{-1} \text d\tilde r^2+\tilde r^{2+\epsilon}\text dS_{2}^2\Biggl)\;.
\label{staticbha}
\ee
Now the 
scalar curvatures becomes
\begin{equation}
\tilde R(\epsilon) = \frac{\tilde r^{-3-\epsilon}(4 \tilde r + \tilde r^\epsilon(-\tilde r(2+\epsilon)(2 + 3 \epsilon) + \epsilon(4+ 3 \epsilon))) }{2 r_S^2} \, .\label{scalarc2}
\end{equation}
Replacing $r$ by $r_S\tilde r$ in the scalar curvature (\ref{scalarc1}) the two expressions for the scalar curvatures agree with each other, $R(r)=\tilde R(r_S\tilde r)$,
as it is required from the general invariance of the scalar curvature.

 The  distance that follows from the scalar curvature (\ref{scalarc2})
 is now given as 
 \begin{equation}
\tilde \Delta_g(\epsilon) \simeq \log \Bigg({R(g^i)\over R(g^f)}\Biggr)(\epsilon)= 2 \log\left(\frac{r_S^f}{r_S^i}\right) \,. 
\end{equation}
Note that for this distance the limit $\epsilon\rightarrow 0$ is trivial.

On the other hand for the  scalar curvature (\ref{scalarc1}) we obtain
\begin{equation}\begin{aligned}
&\Delta_g(\epsilon) \simeq \log \Bigg({R(g^i)\over R(g^f)}\Biggr)(\epsilon) \\
&= \log\left(\frac{r^{-3}(4 (r_S^i)^\epsilon \, r - r^{\epsilon} (r(2+\epsilon)(2 + 3 \epsilon) - r_S^i\epsilon (4+3 \epsilon)))}{r^{-3}(4 (r_S^f)^\epsilon \, r - r^\epsilon (r(2+\epsilon)(2 + 3 \epsilon) - 
r_S^f\, \epsilon (4+3 \epsilon)))}\right).\\
\end{aligned}\end{equation}
If we now take the limit $\epsilon\rightarrow 0$, this expression becomes
\begin{equation}\label{scalarc3}
\Delta_g \simeq   \log\left(\frac{r^{-3}r_S^i -  r^{-2}(2 + \log(r_S^i) -  \log(r))}{r^{-3} r_S^f -  r^{-2}(2 +  \log (r_S^f) -  \log(r))}\right) \,. \\
\end{equation}
We  see that compared to $\tilde \Delta_g$, the distance $\Delta_g$ has an additional dependence on the radial coordinate $r$. Specifically, 
after taking the limit $\epsilon\rightarrow 0$, additional log-terms appear, and
the distance is seemly not invariant under coordinate transformations. This is analogous
to the non-invariance of the metric distance functional, defined in \eqref{dismetr1}.
However for the distances defined via the scalar curvatures,  one can obtain a perfect agreement between $\Delta_g$ and $\tilde \Delta_g$ 
by replacing in (\ref{scalarc3})    the radial coordinate  $r$ in the numerator by $r_S^i$ and in the denominator by $r_S^f$:
\begin{equation}\label{scalarc4}
\Delta_g (r\rightarrow r_S^i,r_S^f)\rightarrow  2 \log\left(\frac{r_S^f}{r_S^i}\right)\equiv\tilde \Delta_g  \,. \\
\end{equation}
This means, that when determining the  distance using the dimensionful coordinate $r$, one has to set $r$ on its horizon values $r_S^i$ or respectively $r_S^f$.
In this way the black hole distance $\Delta_g (r)$ agrees with the one which is computed using the dimensionless coordinate $\tilde r$. 
So the upshot of this discussion is that the correct distance is obtained when measuring the initial and the final values of the distances at the corresponding initial and final horizon values. 
This also should  provide  an explanation why the geometric distance  \eqref{dismetr1} is non-invariant under coordinate transformations, which include the parameter $\tau$.


\section{The metric distance without the volume factor}\label{appvolume}

The distance in metric space can be sometimes encountered without the regulating volume factor of \eqref{dismetr} (see e.g. \cite{DeWitt,gilmedrano1992riemannian}),
\be\label{distanceNoVolume}
\Delta''_g = c \int_{\tau_i}^{\tau_f}  \left[\int_M \sqrt{g} \tr f^2 \right]^{\frac12}\text d\tau\;,
\ee
where again $f= g^{-1}\dot{g}$ while the dot indicates differentiation with respect to the affine parameter $\tau$.
The corresponding geodesic equation reads
\be
\dot f = \tfrac14 \left(\mathrm{tr} f^2 - 2 f \mathrm{tr} f \right) \,.
\label{geodesicsNoVolume}
\ee
The locality of  this equation allows us to formally solve it.
For this purpose we write \eqref{geodesicsNoVolume} explicitly in components
\begin{equation}\label{eq:geodesic2}
\dot f^\mu{}_\nu = \tfrac14 \delta^\mu_\nu f^\kappa{}_\lambda f^\lambda{}_\kappa - \tfrac12 f^\mu{}_\nu f^\kappa{}_\kappa \,,
\end{equation}
and decompose $f^\mu{}_\nu$ into its trace and and a traceless part (we now use the same notation for the scalar $f$ as before for the matrix),
\begin{equation}
f^\mu{}_\nu = \tfrac{1}{d} \delta^\mu_\nu f + \bar f^\mu{}_\nu \qquad \text{with} \qquad \bar f^\mu{}_\mu = 0 \,.
\end{equation}
This allows us to rewrite \eqref{eq:geodesic2} as
\begin{equation}\begin{aligned}\label{eq:geodesic3}
\dot f &= -\tfrac14 f^2 + \tfrac{d}{4} \bar f^\kappa{}_\lambda \bar f^\lambda{}_\kappa \,, \\
\dot {\bar f}{}^\mu{}_\nu &= - \tfrac 12 \bar f^\mu{}_\nu f \,.
\end{aligned}\end{equation}
To proceed we also decompose $\bar f^\mu{}_\lambda \bar f^\lambda{}_\nu$ into its trace and traceless part,
\begin{equation}
\bar f^\mu{}_\lambda \bar f^\lambda{}_\nu = \tfrac1{d^2} \delta^\mu_\nu {\bar f}^2 + \tilde f^\mu{}_\nu \qquad \text{with} \qquad \tilde f^\mu{}_\mu = 0 \,,
\end{equation}
hence we arrive at
\begin{equation}\begin{aligned}
\dot f &= -\tfrac14 \left(f^2 - \bar f^2\right) \,, \\
\dot {\bar f} &= - \tfrac12 \bar f f \,.
\end{aligned}\end{equation}
This is a system of two ordinary, scalar differential equations and has the general solution
\begin{equation}
f(\tau) = \frac{4(\tau - a)}{(\tau - a)^2+b^2} \,,\qquad \bar f(\tau) = \frac{4 b}{(\tau - a)^2+b^2} \,,
\end{equation}
where $a$ and $b$ are space-time dependent integration constants.
We can finally use the second equation of \eqref{eq:geodesic3} to determine also the rest of $\bar f^\mu{}_\nu$ which gives
\begin{equation}
\bar f^\mu{}_\nu = \frac{4}{(\tau - a)^2+b^2} b^\mu{}_\nu \,,
\end{equation}
where we have absorbed an integration constant into the definition of $b^\mu{}_\nu$
and
\begin{equation}
b^\mu{}_\mu = 0 \qquad \text{and} \quad b^\mu{}_\nu b^\nu{}_\mu = \frac{b^2}{d} \,.
\end{equation}
This shows that the most general solution of \eqref{eq:geodesic2} is given by
\begin{equation}\label{eq:fsolution}
f^\mu{}_\nu = \frac{4}{(\tau - a)^2+b^2}\left[\frac{(\tau-a)}{d}\delta^\mu_\nu +b^\mu{}_\nu \right] \,.
\end{equation}
Notice that $a$, $b$ and $b^\mu{}_\nu$ are constant with respect to $\tau$, but can still depend non-trivially on the coordinates of the manifold $M$.
Moreover, the combination
\begin{equation}
f^\mu{}_\nu f^\nu{}_\mu = \frac{16}{d \bigl[ (\tau - a)^2 + b^2 \bigr]} \,,
\end{equation}
which appears in the distance formula \eqref{distanceNoVolume} does not depend on $b^\mu{}_\nu$ but only on $a$ and $b$.
The general solution \eqref{eq:fsolution} for $f^\mu{}_\nu$ can be expressed in terms of $g_{\mu\nu}$ as
\begin{equation}
\dot g_{\mu\nu} = \frac{4}{(\tau - a)^2+b^2}\left[\frac{(\tau-a)}{d}g_{\mu\nu} + g_{\mu\lambda} b^\lambda{}_\nu \right] \,.
\end{equation}
This is a matrix valued ordinary differential equation (pointwise in $M$) and is formally solved by
\begin{equation}
g_{\mu\nu}(\tau) = \left[(\tau - a)^2+b^2\right]^\frac2d g^0_{\mu\lambda} \exp\left[\frac{4 b^\lambda{}_\nu}{b} \arctan \left(\frac{\tau-a}{b}\right) \right] \,,
\end{equation}
where the exponential is to be understood as the matrix exponential of $b^\mu{}_\nu$.

To illustrate that \eqref{distanceNoVolume} produces different results than the distance formulae \eqref{dismetr} with varying normalization factor $V_M^{-1}$, we can for example consider the special case
\begin{equation}
a = \mathrm{const}. \qquad\text{and}\qquad b = b^\mu{}_\nu = 0  \,.
\end{equation}
This corresponds to the conformal rescaling
\begin{equation}\label{eq:conformalgeodesic}
g_{\mu\nu}(\tau) = \alpha(\tau) g^0_{\mu\lambda} \,,\qquad \alpha(\tau) = (\tau - a)^\frac4d \,.
\end{equation}
Notice, that this is the same path as \eqref{conformalAnsatz} but parametrized differently, i.e.~here \eqref{dismetr} and \eqref{distanceNoVolume} give rise to different proper times.
Moreover,
we can insert \eqref{eq:conformalgeodesic} into \eqref{distanceNoVolume} and obtain
\be
\Delta''_g=c\left(\int_{\cal M}\sqrt{g^0}\right)^{1/2}\frac{4}{\sqrt{d}}\left(\alpha_f^{\frac{d}{4}}-\alpha_i^{\frac{d}{4}}\right) \ .
\ee
For AdS space-times, it was argued in \cite{Lust:2019zwm} that matching the scaling of the mass of (Kaluza-Klein) towers of states suggests a logarithmic behaviour of the distance with respect to the cosmological constant. Thus, since \eqref{dismetr} reproduces such a logarithmic scaling whereas \eqref{distanceNoVolume} does not, the former is preferred in this paper. 
On the other hand, the geodesic equation \eqref{geodesicsNoVolume} is local, unlike \eqref{geodesicsVolume}. Some properties of the former can be found in \cite{gilmedrano1992riemannian} and references therein.

\section{Some more details about black hole, microstates, the N-portrait  and Goldstone modes}\label{appC}

The following discussion is largely based on the black hole N-portrait model, which was developed in \cite{portrait,portrait1,gold,Dvali:2015rea} (see also \cite{Dvali:2016lnb}).
We consider a black hole (e.g.~Schwarzschild black hole) of mass M and entropy ${\cal S}=M^2$. In the following we denote the entropy ${\cal S}$
by $N$, i.e.~$M=N^{1/2}$. The number of black hole microstates $n$ is then of the order $n=\exp(N)$.
Classically and for large masses, i.e. $N\rightarrow \infty$, the number of microstates grows exponentially and for infinite $N$ they are all degenerate.
However in the quantum case for finite $N$, it is very natural to assume that the degeneracy among the microstates is lifted.
Then the energy eigenvalues among the microstates are distributed within some finite energy interval, which we will denote by $\epsilon^{max}$.
Since there are $n=\exp(N)$ microstates, it follows that the average energy differences between the microstates are of the order 
\begin{equation}
\Delta E^{average}= \exp(-N)\epsilon^{max}\, .
\end{equation}
In the following we like to argue that, although  $\Delta E^{average}$ is exponentially suppressed, 
there are only $N+1$ different energy levels among the microstates, and that
the actual energy differences among the microstates are only suppressed by $1/N$ compared to 
$\Delta E^{average}$, meaning that many of the microstates must be still degenerate.
Moreover there will be $N$ collective Goldstone modes that interpolate between the different microstate energy levels.
These light Goldstone modes describe the transitions between the different black hole microstates and their energies are of the same order as the differences
between the energy levels of the black hole microstates.

Concretely we assume the the Hilbert space $H_{bh}$ of a black hole is the direct product of $N$ identical copies of sub-Hilbert spaces $H_n$:
\begin{equation}
H_{bh}=\prod_{n=1}^N H_n\, .
\end{equation}
E.g. $H_n$ can be the Hilbert space of a CFT with $K3$ target space, as it is the case for the dual description of certain extremal black holes \cite{Strominger:1996sh}.
Here we assume for simplicity that $dim(H_n)=2$, and hence $H_n$ contains two states $|\uparrow\rangle$ and $|\downarrow\rangle$, also denoted by $|\pm\rangle$.
The corresponding energy eigenvalues of $|\pm\rangle$ in $H_n$ are
\begin{equation}
E^\pm_n=E_0\pm \Delta E/2\, .
\end{equation}

A general black hole microstate in $H_{bh}$ has now the form
\begin{equation}
|\Psi\rangle_{bh}=|\pm,\pm ,\dots,\pm\rangle\, .
\end{equation}
It follows that the dimension of $H_{bh}$ is $dim(H_{bh})=2^N$, which is of order $\exp(N)$.\footnote{Actually for a black hole, the microstates should be highly entangled states.}
The energy (mass) of a black holes microstate is of the order
\begin{equation}
E_{bh}=\sum_{n=1}^nE^\pm_n\, .
\end{equation}
We see that we have $N+1$ different energy levels with $N+1$ different energy eigenvalues
\begin{equation}\label{bheigenvalues}
E_{bh}^i=NE_0+\biggl({N\over 2}-i\biggr)\Delta E\, ,\qquad i=0,\dots, N\, .
\end{equation}
The degeneracy of a state at the energy level $E^{bh}_i$ can be also easily determined, it is of the order ${N}\choose{i} $.
In order that the  black hole microstates with energy eigenvalues as given in (\ref{bheigenvalues}) account for the correct overall black hole mass, we have to require that
\begin{equation}
NE_0=\sqrt N\, ,\qquad \Longrightarrow\quad E_0=1/\sqrt N\, .
\end{equation}

Furthermore we see that we have $N$ collective Goldstone modes that describe the light (soft) excitations between the black hole microstates. The $N$ different energies $\epsilon_j$
of the Goldstone modes are determined by the energy differences between the black hole microstates and are therefore given as follows:
\begin{equation}\label{gbenergies}
\epsilon_j=j\Delta E\, ,\qquad j=1,\dots, N\, .
\end{equation}
It follows that total size of the energy gap is given as
\begin{equation}\label{bheigenvalues0}
\epsilon^{max}=N\Delta E\, ,
\end{equation}
and hence for a given size of the energy gap the energy difference scale as $\Delta E=\epsilon^{max}/N$.

\vskip0.3cm
We can now consider a few different reasonable cases for the choice of $\epsilon^{max}$ (all in Planck units):

\begin{itemize}
\item {\sl Planck mass gap}

Here the energy gap is of the order of the Planck mass:
\begin{equation}\label{bheigenvalues1}
\epsilon^{max}=1\,, \quad \Longrightarrow\quad \Delta E= {1\over N}\, .
\end{equation}

\item {\sl Hawking temperature gap}

Here the gap is of the order of the inverse horizon size $1/r_S$, which is of the order of the Hawking temperature:
\begin{equation}\label{bheigenvalues2}
\epsilon^{max}=T_H={1/\sqrt N}\, ,\quad \Longrightarrow\quad \Delta E= N^{-3/2}\, .
\end{equation}

\item
 {\sl Black hole N-portrait  gap}

 Due to quantum criticality of the graviton (BEC) condensate there is in additional suppression factor of $1/N$ compared to the previous case \cite{portrait,portrait1,gold,Dvali:2015rea}:
 \begin{equation}\label{bheigenvalues3}
\epsilon^{max}={1/(N\sqrt N})\, ,\quad \Longrightarrow\quad \Delta E= N^{-5/2}\, .
\end{equation}
Moreover, due to the spherical symmetry of the black hole, the equally spaced Goldstone spectrum in (\ref{gbenergies}), is replaced by the 
 following Goldstone spectrum   \cite{Averin:2016ybl,Averin:2016hhm}:
 \begin{equation}\label{gbenergies1}
\epsilon_l=l^2\Delta E\, ,\qquad l=1,\dots, \sqrt N\, .
\end{equation}
 The quantum number $l$ can be viewed as an angular momentum number, and hence each Goldstone mode has an $2l+1$ degeneracy such that there in total again
 $\sum_{l=1}^{\sqrt N}(2l+1)=N$ Goldstone modes.

\end{itemize}

As explained in \cite{Averin:2016ybl,Averin:2016hhm}
the N Goldstone modes with energies, as given in (\ref{gbenergies})  and  (\ref{gbenergies1}), are the pseudo Goldstone bosons of the supertranslations symmetries (i.e. BMS like symmetries) at the black hole horizon. 
They correspond to the so-called soft black hole hair \cite{Hawking:2016msc}.
The associated charges $Q^{BMS}$ transform one black hole microstate into another one:
\begin{equation}
|\Psi'\rangle_{bh}=Q^{BMS}|\Psi\rangle_{bh}.
\end{equation}
For large $N\rightarrow\infty$ the Goldston modes become massless and there are infinitely many  BMS supertranslations.
Since the black hole microstate  $|\Psi\rangle_{bh}$
is not invariant under the action of $Q^{BMS}$, it  follows that the supertranslations at the horizon are spontaneously broken in the black hole vacuum.
Actually for $N=\infty$ the Goldstone modes are exactly massless, and the (interior) black hole geometry becomes flat Minkowski space with the horizon pushed to null infinity.
In this case the supertranslations are just the standard BMS transformations and the infinite entropy $N=\infty$ describes the infinite degeneracy, i.e. entropy  of flat space space 
under BMS transformations    \cite{Dvali:2015rea}.  However for finite $N$ the supertranslation symmetries are explicitly broken, and hence the N Goldstone acquire a finite mass.

\end{appendices}


\begin{thebibliography}{99}
\bibitem{Perelman:2006un}
  G.~Perelman,
  ``The Entropy formula for the Ricci flow and its geometric applications,''
  math/0211159 [math-dg].
\end{thebibliography}
\end{document}